\documentclass[10pt,a4paper]{article}

\usepackage{fullpage,enumerate}
\usepackage{amssymb,amsmath,mathrsfs}
\usepackage{xcolor}

\newtheorem{definition}{Definition}[section]
\newtheorem{lemma}[definition]{Lemma}
\newtheorem{proposition}[definition]{Proposition}
\newtheorem{theorem}[definition]{Theorem}

\newtheorem{remark}[definition]{Remark}

\newenvironment{proof*}{\smallskip\par\noindent\emph{Proof. }
 \ignorespaces}{\hfill$\Box$\smallskip\par\ignorespaces}
\newenvironment{proofsketch*}{\smallskip\par\noindent
 \emph{Sketch of proof. }\ignorespaces}
 {\hfill$\oslash$\smallskip\par\ignorespaces}

\newcommand{\N}{\ensuremath{\mathbb{N}}}
\newcommand{\R}{\ensuremath{\mathbb{R}}}
\newcommand{\C}{\ensuremath{\mathbb{C}}}

\newcommand{\A}{\ensuremath{\mathcal{A}}}
\newcommand{\cH}{\ensuremath{\mathcal{H}}}

\title{\textbf{Stress tensor bounds on quantum fields}}
\author{Ko Sanders\thanks{E-mail:
jacobus.sanders@fau.de}}
\date{Department Mathematik, FAU
Erlangen-N\"urnberg, Cauerstra{\ss}e 11, 91058 Erlangen\\[2ex]
17 April 2024}

\begin{document}

\maketitle

\section*{Abstract}

The singular behaviour of quantum fields in Minkowski space can often be bounded by polynomials of the Hamiltonian $H$. These so-called $H$-bounds and related techniques allow us to handle pointwise quantum fields and their operator product expansions in a mathematically rigorous way. A drawback of this approach, however, is that the Hamiltonian is a global rather than a local operator and, moreover, it is not defined in generic curved spacetimes. In order to overcome this drawback we investigate the possibility of replacing $H$ by a component of the stress tensor, essentially an energy density, to obtain analogous bounds. For definiteness we consider a massive, minimally coupled free Hermitean scalar field. Using novel results on distributions of positive type we show that in any globally hyperbolic Lorentzian manifold $M$ for any $f,F\in C_0^{\infty}(M)$ with $F\equiv 1$ on $\mathrm{supp}(f)$ and any timelike smooth vector field $t^{\mu}$ we can find constants $c,C>0$ such that $\omega(\phi(f)^*\phi(f))\le C(\omega(T^{\mathrm{ren}}_{\mu\nu}(t^{\mu}t^{\nu}F^2))+c)$ for all (not necessarily quasi-free) Hadamard states $\omega$. This is essentially a new type of quantum energy inequality that entails a stress tensor bound on the smeared quantum field. In $1+1$ dimensions we also establish a bound on the pointwise quantum field, namely $|\omega(\phi(x))|\le C(\omega(T^{\mathrm{ren}}_{\mu\nu}(t^{\mu}t^{\nu}F^2))+c)$, where $F\equiv 1$ near $x$.

\section{Introduction}\label{sec:intro}

Quantum fields are typically very singular objects. Not only must we treat $\phi(x)$ as an operator-valued distribution \cite{Wightman1996}, rather than an operator-valued function, but for bosonic fields the averaged expression $\phi(f)=\int \phi(x)f(x)\mathrm{d}x$ with some test function $f$ normally yields an unbounded operator, so its action is not defined on all vectors in a Hilbert space.

In Minkowski space and in the vacuum representation, this singular behaviour can typically be controlled using bounds on the quantum field in terms of the Hamiltonian operator $H$. Indeed, under quite general conditions there hold \emph{polynomial $H$-bounds}, i.e.
expressions of the form $\phi(f)(1+H)^{-k}$ determine bounded operators for all sufficiently large $k$, cf. \cite{FH1981} and Sec.14.3 of \cite{BW1992}. Moreover, even pointlike quantum fields can be handled using the boundedness of quadratic forms of the form $(1+H)^{-k}\phi(x)(1+H)^{-k}$ for sufficiently large $k$. The use of $H$-bounds for theories with sufficiently good phase space behaviour gives rise to operator product expansions, that are a useful tool to investigate interacting QFTs \cite{Bos2005a,Bos2005b}.

One major drawback of these $H$-bounds is that the operator $H$ is of a global nature and only well-defined on stationary Lorentzian manifolds. I.e., for quantum fields in a general curved spacetime the singular behaviour of quantum field cannot be controlled by an $H$-bound in this way. This is a major obstruction for attempts to extend structural results about pointlike fields and operator product expansations to a generally covariant setting.

In this paper we will investigate whether this drawback can be overcome by replacing the Hamiltonian operator $H$ by an average of a suitable component of the quantum stress tensor, essentially an averaged energy density. For definiteness we will consider a massive minimally coupled free scalar field, for which the quantum stress tensor is well understood, including its renormalization and all remaining renormalization freedom \cite{HW2005}. In this case the stress tensor is a local and covariant quantum field, so replacing $H$ by a component of the stress tensor should give rise to bounds that behave in a local and covariant way. Our emphasis, however, will be on a single Lorentzian manifold and for convenience we will renormalize the stress tensor by subtracting a reference Hadamard two-point distribution, rather than a local and covariant Hadamard parametrix.

Using an apparently new mathematical result on distributions of positive type (as defined in Section \ref{sec:positivity} below), we will establish a polynomial bound on the smeared quantum field $\phi(f)$ in terms of the averaged stress tensor $T^{\mathrm{ren}}_{\mu\nu}(t^{\mu}t^{\nu}F^2)$, where $f,F$ are test functions on a globally hyperbolic Lorentzian manifold $M$ such that $F\equiv 1$ on $\mathrm{supp}(f)$ and $t^{\mu}$ is any smooth time-like vector field. This stress tensor bound essentially takes the form of a new kind of \emph{quantum energy inequality} (QEI) \cite{Few2017,KS2020},
\begin{align}
\omega(\phi(f)^*\phi(f))&\le C(\omega(T^{\mathrm{ren}}_{\mu\nu}(t^{\mu}t^{\nu}F^2))+c)\notag
\end{align}
for some constants $c,C>0$ and all (not necessarily quasi-free) Hadamard states $\omega$. Here the constants $c$ and $C$ are independent of the state $\omega$. In the GNS-representation of any Hadamard state this bound leads to the analogue of an $H$-bound, where $H+1$ is replaced by the operator
$T^{\mathrm{ren}}_{\mu\nu}(t^{\mu}t^{\nu}F^2)+c$ (or rather by the Friedrichs extension of its representative). Analogous local energy bounds have been studied for chiral CFTs in $1+1$ dimensions, cf. \cite{CTW2022} and references cited therein.

In $1+1$ dimensions we will also establish a bound on the pointwise quantum field, namely
\begin{align}
|\omega(\phi(x))|&\le C(\omega(T^{\mathrm{ren}}_{\mu\nu}(t^{\mu}t^{\nu}F^2))+c)\,,\notag
\end{align}
where $F\equiv 1$ in a neighbourhood of the point $x$ and the constants $c,C>0$ may be different than before. We conjecture that a similar result involving higher powers of the stress tensor is valid in higher dimensions.

This paper is organised as follows. Section \ref{sec:positivity} describes the mathematical results on distributions of positive type that we need. It uses some results about Sobolev wave front sets, which are relegated to Appendix \ref{app:WFs}. Section \ref{sec:smearedbounds} establishes the stress tensor bound for the smeared massive free scalar field. The stress tensor bound for pointwise fields in $1+1$ dimensions is contained in Section \ref{sec:pointwisebound}. We conclude with a brief discussion in Section \ref{sec:discussion}.

\section{On distributions of positive type}\label{sec:positivity}

In this section we will prove some results on distributions of positive type that we will need later in Section \ref{sec:smearedbounds}. We will prove these in the following general setting. Let $M$ be a smooth manifold of dimension $d\in\N$ with a smooth volume form $\mu$. A distribution $u$ on $M\times M$ is said to be \emph{of positive type} when $u(\mu\bar{f}\otimes \mu f)\ge0$ for all $f\in C_0^{\infty}(M)$.\footnote{Throughout this paper, test functions are allowed to be complex-valued, unless indicated otherwise.} We will assume that $M$ is orientable and $\mu$ is nowhere vanishing.

In the following theorem, $\mathcal{Z}$ denotes the zero section of the bundle $T^*M^{\times 2}$.
\begin{theorem}\label{thm:positivity}
Let $u_1,u_2$ be distributions on $M^{\times 2}$ of positive type such that for all $(x,\xi)\in T^*M^{\times 2}\setminus\mathcal{Z}$ there are $s_1,s_2\in \R$ with $s_1+s_2\ge0$, $(x,\xi)\not\in WF^{(s_1)}(u_1)$ and $(x,-\xi)\not\in WF^{(s_2)}(u_2)$. Then $u_1\cdot u_2$ is also of positive type.
\end{theorem}
\begin{proof*}
A product of distributions is not always well-defined in a natural way. \cite{Obe1986} considers the product of distributions in $\R^n$ as defined by Ambrose under suitable hypotheses and compares it with several alternative definitions. In particular, Cor.~3.1 loc.cit. shows that the Sobolev wave front set condition on $u_1$ and $u_2$ ensures that the product $u_1u_2$ is a well-defined distribution. This result can be generalised to distributions on the smooth manifold $M$ using local charts and a partition of unity argument.

We fix an arbitrary $f\in C_0^{\infty}(M)$ and we choose $F\in C_0^{\infty}(M,\R)$ such that $F\equiv 1$ on $\mathrm{supp}(f)$. The distributions
$u_1':=(F\otimes F)\cdot u_1$ and $u_2':=(\bar{f}\otimes f)\cdot u_2$ are again of positive type and $(u_1\cdot u_2)(\mu\bar{f}\otimes \mu f)=u_1'((\mu\otimes\mu)\cdot u_2')$. The $u_i'$ are also compactly supported and they satisfy the same wave front set estimate as the $u_i$. Replacing $u_i$ by $u_i'$ it then suffices to show that $u_1((\mu\otimes\mu)\cdot u_2)\ge0$, where we may assume that both $u_i$ have compact support.

Let $B\subset\R^d$ denote the open unit ball. Because $F$ has compact support, we can choose a finite number of charts $\kappa_k:O_k\to B$, $k=1,\ldots,n$ such that the union of the domains $O_j\subset M$ covers the support of $F$. Furthermore, we can choose $\chi_k\in C_0^{\infty}(O_k,\R)$ such that
$\sum_{k=1}^n\chi_k^2\equiv 1$ on $\mathrm{supp}(F)$. For all $k,l\in\{1,\ldots,n\}$ we define the distributions $v_{1,kl}:=(\kappa_k\times\kappa_l)_*((\chi_k\otimes\chi_l)\cdot u_1)$ and the distribution densities
$v_{2,kl}:=(\kappa_k\times\kappa_l)_*((\chi_k\mu\otimes\chi_l\mu)\cdot u_2)$ in $\R^{2d}$, which are supported in $B^{\times 2}$. Using
$\sum_{k=1}^n\chi_k^2\equiv 1$ we then have $u_1((\mu\otimes\mu)\cdot u_2)=\sum_{k,l=1}^nv_{1,kl}(v_{2,kl})$ and the positive type property of the $u_i$ now means that
\begin{align}
\sum_{k,l=1}^nv_{i,kl}(\bar{f}_k\otimes f_l)&\ge0\label{eqn:posmatrix}
\end{align}
for all $f_1,\ldots,f_n\in C_0^{\infty}(\R^d)$. For the distribution densities $v_{2,kl}$ this follows immediately from the definitions. For $v_{1,kl}$ we use the fact that we can write $f_k=\mu_kg_k$ with $g_k\in C_0^{\infty}(B)$ and $\mu_k$ such that $(\kappa_k)_*(\mu)=\mu_k\mathrm{d}^dx$.

Now let $\eta\in C_0^{\infty}(\R^d)$ with $\eta\ge0$ and $\int\eta=1$. For each $\lambda>0$ we let $\eta_{\lambda}(x)=\lambda^{-d}\eta(\lambda^{-1}x)$ 
and $\tilde{\eta}_{\lambda}(x)=\eta_{\lambda}(-x)$ and we define the convolutions $v_{i,kl}^{(\lambda)}:=(\eta_{\lambda}\otimes\eta_{\lambda})*v_{i,kl}$. Note that the $v_{i,kl}^{(\lambda)}$ are compactly supported and smooth and they also satisfy (\ref{eqn:posmatrix}). Indeed, because the functions $v_{i,kl}^{(\lambda)}$ are smooth, we even find
\begin{align}
\sum_{k,l=1}^nv_{i,kl}^{(\lambda)}(\bar{f}_k\otimes f_l)&=\sum_{k,l=1}^nv_{i,kl}((\tilde{\eta}_{\lambda}*\bar{f}_k)\otimes (\tilde{\eta}_{\lambda}*f_l))
\ge0\label{eqn:posmatrix2}
\end{align}
for all distributions $f_1,\ldots,f_n$ on $\R^d$. For any $\lambda>0$ we may view the matrix of functions $\left(v_{2,kl}^{(\lambda)}\right)_{k,l=1}^n$ as the integral kernel of a linear operator $X$ on $L^2(\R^d,\C^n)$, defined by $(Xf)_k=\sum_{l=1}^nv_{2,kl}^{(\lambda)}f_l$, where $f=(f_1,\ldots,f_l)\in L^2(\R^d,\C^n)$. Because $(v_{2,kl}^{(\lambda)})_{k,l=1}^n\in L^2(\R^{2d},\C^{n^2})$ the operator $X$ is Hilbert-Schmidt (cf. \cite{KR1997} Prop.2.6.9). 
Equation (\ref{eqn:posmatrix2}) shows that $X$ is also a positive operator, so we may write 
\begin{align}
X&=\sum_{j=1}^{\infty}x_j\psi_j\otimes\psi_j^*\notag
\end{align}
for some orthonormal basis $\{\psi_j\}_{j\in\N}$ of $L^2(\R^d,\C^n)$ and $x_j\ge0$ (cf. \cite{RS1980}, Thm.VI.16). Note that the series
\begin{align}
v_{2,kl}^{(\lambda)}(x,y)&=\sum_{j=1}^{\infty}x_j \psi_{j,k}(x)\overline{\psi_{j,l}(y)}\notag
\end{align}
converges in $L^2(\R^{2d})$. Because also $v_{1,kl}^{(\rho)}\in L^2(\R^{2d})$ for all $\rho>0$ we then have
\begin{align}
\sum_{k,l=1}^nv_{1,kl}^{(\rho)}(v_{2,kl}^{(\lambda)})&=
\sum_{j=1}^{\infty}x_j\sum_{k,l=1}^nv_{1,kl}^{(\rho)}(\psi_{j,k}\otimes \overline{\psi_{j,l}})\ge0\label{eqn:posmatrix3}
\end{align}
by Equation (\ref{eqn:posmatrix2}).

The comparison of various notions of products of distributions in \cite{Obe1986} shows that the Sobolev wave front set condition also implies
\begin{align}
v_{1,kl}(v_{2,kl})&=\lim_{\lambda\to0^+}v_{1,kl}^{(\lambda)}(v_{2,kl}^{(\lambda)})\notag
\end{align}
and hence
\begin{align}
u_1((\mu\otimes\mu)\cdot u_2)&=\sum_{k,l=1}^nv_{1,kl}(v_{2,kl})=\lim_{\lambda\to0^+}\sum_{k,l=1}^nv_{1,kl}^{(\lambda)}(v_{2,kl}^{(\lambda)})\ge0\notag
\end{align}
by (\ref{eqn:posmatrix3}). This completes the proof.
\end{proof*}

Next we want to give a construction of a certain distribution of positive type that we will need in Section \ref{sec:smearedbounds}. We will use the following lemma.
\begin{lemma}\label{lem:v0}
For $l\in\N$ let $v\in\mathcal{D}'(\R)$ such that
\begin{align}
\hat{v}(k)&:=\left\{
\begin{array}{ll}
\frac12(1+k^2)^{-l}&\mathrm{if}\ k>0\\
1-\frac12(1+k^2)^{-l}&\mathrm{if}\ k\le0
\end{array}
\right.\,.\notag
\end{align}
Then $WF^{(s)}(v)\subset \{0\}\times\R_{<0}$ for all $s<2l-\frac12$.
\end{lemma}
\begin{proof*}
For $u\in\mathcal{D}'(\R)$ with $\hat{u}(k)=(1+k^2)^{-l}$ we note that $u$ is smooth on $\R\setminus\{0\}$ and $WF^{(s)}(u)=\emptyset$ iff
$s-2l<-\frac12$ (cf. Cor.8.4.7. in \cite{HormanderNlin}). Because $u$ is real and even it then follows that $WF^{(s)}(u)=\{0\}\times(\R\setminus\{0\})$ iff $s-2l\ge-\frac12$. Now define $u_{\pm}$ by $\widehat{u_{\pm}}(k)=\hat{u}(k)\theta(\pm k)$, where $\theta$ is the Heaviside function. Because $\widehat{u_+}$ is supported on $k>0$ we must have $WF^{(s)}(u_+)\subset\R\times\R_{>0}$ for all $s\in\R$. Similarly, $WF^{(s)}(u_-)\subset\R\times\R_{<0}$ and hence $WF^{(s)}(u)=WF^{(s)}(u_+)\cup WF^{(s)}(u_-)$ as a disjoint union, so $WF^{(s)}(u_+)=\{0\}\times\R_{>0}$ iff $s\ge 2l-\frac12$. Analogously, defining $v_{\pm}$ such that $\widehat{v_{\pm}}(k)=\hat{v}(k)\theta(\pm k)$, we also have $WF^{(s)}(v)=WF^{(s)}(v_+)\cup WF^{(s)}(v_-)$ as a disjoint union. $v_+=\frac12u_+$ proves the result. (We only require an estimate on the part of the wave front set contained in
$\R\times\R_{>0}$, so we don't need to determine $WF^{(s)}(v_-)$ for all $s$.)
\end{proof*}

The construction in the lemma below makes use of the delta distribution $\delta_{\mu}$ supported on the diagonal of $M^{\times 2}$, defined by $\delta_{\mu}(\mu g, \mu h):=\int gh\mu$ for all $g,h\in C_0^{\infty}(M)$.

\begin{lemma}\label{lem:positiveu}
Assume $d\ge2$ and let $f,F\in C_0^{\infty}(M)$ such that $F$ is real-valued and $F>0$ on $\mathrm{supp}(f)$. Let $s\in\R$ and $t$ a smooth vector field on $M$ such that $t\not=0$ on $\mathrm{supp}(F)$. Then there is a compactly supported distribution $u$ on $M^{\times 2}$ and a constant $C>0$ such that $Cu(x,y)-f(x)\overline{f(y)}$ is of positive type, $u(x,y)+u(y,x)=F(x)F(y)\delta_{\mu}(x,y)$ and $WF^{(s)}(u)\subset \{((x,y),(p,q))\in T^*M^{\times 2}|\ 
p(t)\le0\le q(t)\}$.
\end{lemma}
\begin{proof*}
We can cover $\mathrm{supp}(F)$ by a finite set of open coordinate charts $\{(O_j,\kappa_j)\}_{j=1}^n$ on which the vector field $t$ is the first coordinate derivative $t=\partial_{x_1}$. We then have $\mu=\mu_j\kappa_j^*\mathrm{d}^dx$ for a smooth non-vanishing function $\mu_j$. Because $d\ge2$ we can 
change the sign of $x_2$ if necessary to ensure that $\mu_j>0$.

We may choose $\chi_j\in C_0^{\infty}(O_j)$ such that $\chi_j\ge0$ and $\sum_{j=1}^n\chi_j^2\equiv1$ on $\mathrm{supp}(F)$. We then define the distribution $u$ on $M^{\times 2}$ by setting
\begin{align}
u(x,y)&:=(2\pi)^{-d}\int\mathrm{d}^dk\sum_{j=1}^n F(x)\chi_j(x) \mu_j(x)^{-\frac12} e^{i\kappa_j(x)\cdot k} F(y)\chi_j(y)\mu_j(y)^{-\frac12}
e^{-i\kappa_j(y)\cdot k}\hat{v}(k_1)\,,\notag
\end{align}
with $v$ as in Lemma \ref{lem:v0} for some $l>\frac{s}{2}+\frac{d}{4}$. Note that $u(x,y)$ is well-defined, because $\chi_j=0$ outside the coordinate neighbourhood $O_j$. It is also compactly supported and because $\hat{v}(k_1)+\hat{v}(-k_1)=1$ we have
\begin{align}
u(x,y)+u(y,x)&=\sum_{j=1}^n F(x)\chi_j(x) F(y)\chi_j(y)\delta_{\mu}(x,y)=F(x)F(y)\delta_{\mu}(x,y)\,,\notag
\end{align}
where the factors of $\mu_k$ in the definition of $u$ are needed to produce the correct measure for $\delta_{\mu}$.

To show the wave front set estimate we consider the distribution $u_0$ on $\R^{2d}$ defined by
$u_0(x,y)=(2\pi)^{-d}\int\mathrm{d}^dk e^{i(x-y)\cdot k}\hat{v}(k_1)$. Using the linear change of coordinates $(X,Y)=(x-y,x+y)$, which acts on covectors as
$(P,Q)=\left(\frac{p-q}{2},\frac{p+q}{2}\right)$, we find that $u_0(x,y)=v(X_1)\otimes \delta(X')\otimes 1(Y)$, where $X=(X_1,X')$. Hence, $((x,y),(p,q))\in WF^{(s)}(u_0)$ exactly when $((X,Y),(P,Q))\in WF^{(s)}(v\otimes\delta\otimes 1)$, which implies $Q=0$ and $(X,P)\in WF^{(s)}(v\otimes\delta)$ by Lemma \ref{lem:tensorWFs}. By our choice of $l$ we can combine Lemma \ref{lem:v} and Lemma \ref{lem:v0} to see that $P_1\le0$ and $X'=0$. We therefore find
\begin{align}
WF^{(s)}(u_0)&\subseteq\{((x,y),(p,-p))|\ x'=y', p_1\le0\}\,.\notag
\end{align}
Because $p_1\le0$ exactly when $p(t)\le0$ we must also have $WF^{(s)}(u)\subseteq\{((x,y),(p,q))|\ p_1(t)\le0\le q_1(t)\}$.

It remains to show that $Cu(x,y)-f(x)\overline{f(y)}$ is of positive type for a suitable $C>0$. For this purpose we define $\tilde{f}\in C_0^{\infty}(M)$ by
$\tilde{f}(x)=\frac{f(x)}{F(x)}$ when $F(x)\not=0$ and $\tilde{f}(x)=0$ otherwise, so that $f=\tilde{f}F$. For any $g\in C_0^{\infty}(M)$ we can estimate
\begin{align}
|\langle f,g\rangle_{\mu}|^2&=\left|\sum_{j=1}^n1\cdot\langle \chi_j^2\tilde{f},Fg\rangle_{\mu}\right|^2
\le n\sum_{j=1}^n|\langle\chi_j^2\tilde{f},Fg\rangle_{\mu}|^2=n\sum_{j=1}^n|\langle\chi_j\tilde{f},\chi_jFg\rangle_{\mu}|^2\notag
\end{align}
using $\sum_{j=1}^n\chi_j^2\equiv1$, the Cauchy-Schwarz inequality and the inner product $\langle,\rangle_{\mu}$ of $L_2(M,\mu)$. Using the charts
$\kappa_j$ and denoting the usual $L_2$ inner product on $\kappa_j(O_j)$ by $\langle,\rangle$ we can rewrite this estimate as
\begin{align}
|\langle f,g\rangle_{\mu}|^2&\le n\sum_{j=1}^n|\langle(\kappa_j)_*(\mu_j^{\frac12}\chi_j\tilde{f}),(\kappa_j)_*(\mu_j^{\frac12}\chi_jFg)\rangle|^2\notag\\
&=\frac{n}{(2\pi)^{2d}}\sum_{j=1}^n\left|\int \overline{[(\kappa_j)_*(\mu_j^{\frac12}\chi_j\tilde{f})]\hat{\ }(k)}\cdot
[(\kappa_j)_*(\mu_j^{\frac12}\chi_jFg)]\hat{\ }(k)\mathrm{d}^dk\right|^2\notag\\
&=\frac{n}{(2\pi)^{2d}}\sum_{j=1}^n\left|\int \overline{(1+k_1^2)^{\frac{l}{2}}[(\kappa_j)_*(\mu_j^{\frac12}\chi_j\tilde{f})]\hat{\ }(k)}\cdot
(1+k_1^2)^{\frac{-l}{2}}[(\kappa_j)_*(\mu_j^{\frac12}\chi_jFg)]\hat{\ }(k)\mathrm{d}^dk\right|^2\,,\notag
\end{align}
where $[X]\hat{\ }$ denotes the Fourier transform of $X$. Another application of the Cauchy-Schwarz inequality and the estimate
$\hat{v}(k_1)\ge\frac12(1+k_1^2)^{-l}$ (cf. Lemma \ref{lem:v0}) then yield
\begin{align}
|\langle f,g\rangle_{\mu}|^2&\le\frac{n}{(2\pi)^{2d}}\sum_{j=1}^n\int (1+k_1^2)^l\left|[(\kappa_j)_*(\mu_j^{\frac12}\chi_j\tilde{f})]\hat{\ }(k)\right|^2\mathrm{d}^dk\cdot \int (1+k_1^2)^{-l}\left|[(\kappa_j)_*(\mu_j^{\frac12}\chi_jFg)]\hat{\ }(k)\right|^2\mathrm{d}^dk\notag\\
&\le\frac{2n}{(2\pi)^{2d}}\sum_{j=1}^n\int (1+k_1^2)^l\left|[(\kappa_j)_*(\mu_j^{\frac12}\chi_j\tilde{f})]\hat{\ }(k)\right|^2\mathrm{d}^dk
\cdot \int \hat{v}(k_1)\left|[(\kappa_j)_*(\mu_j^{\frac12}\chi_jFg)]\hat{\ }(k)\right|^2\mathrm{d}^dk\notag\\
&\le Cu(\mu g,\mu g)\,,\notag
\end{align}
where we used the definition of $u$ and we introduced
\begin{align}
C&:=\frac{2n}{(2\pi)^d}\max_{j=1,\ldots,n}\int (1+k_1^2)^l\left|[(\kappa_j)_*(\mu_j^{\frac12}\chi_j\tilde{f})]\hat{\ }(k)\right|^2\mathrm{d}^dk\,.\notag
\end{align}
This proves that $f(x)\overline{f(y)}\le Cu(x,y)$, as desired.
\end{proof*}

\begin{remark}
When $d=1$ the analogous conclusion of Lemma \ref{lem:positiveu} still holds if we additionally assume that $\mu(t)>0$, because this ensures that the $\mu_j$ are still positive functions. if $\mu(t)<0$ we could replace $t$ by $-t$, which leads to a change in the estimate for $WF^{(s)}(u)$.
\end{remark}

\section{Stress tensor bounds on smeared free scalar fields}\label{sec:smearedbounds}

Consider a smooth globally hyperbolic Lorentzian manifold $(M,g)$ with dimension $d=n+1\ge2$. We assume that $M$ is time-oriented and we consider a massive, minimally coupled real linear scalar field of mass $m>0$ on $M$. I.e., we consider the Klein-Gordon equation $P\varphi=0$, where $P=\Box+m^2$ and we let $E^-$, respectively $E^+$, denote the advanced, respectively retarded, fundamental solutions for $P$, viewed as distributions on $M^{\times 2}$. A classical field configuration is a solution $\varphi$ of this equation. Its stress tensor is given by
\begin{align}
T_{\mu\nu}[\varphi]&=\nabla_{\mu}\varphi\cdot \nabla_{\nu}\varphi-\frac12g_{\mu\nu}(|\nabla\varphi|^2-m^2\varphi^2)\,.\label{eq:Tclassical}
\end{align}

To be specific, we describe the quantized theory by the Weyl-algebra $\A$, which is a complex $^*$-algebra generated by operators $W(f)$, $f\in C_0^{\infty}(M,\R)$, subject to the relations
\begin{enumerate}[(i)]
\item $W(f)^*=W(-f)$,
\item $W(Pf)=W(0)$,
\item $W(f)W(h)=e^{-i\frac12E(f,h)}W(f+h)$,
\end{enumerate}
where $E=E^--E^+$. Note that $W(0)=1$ is the identity operator and each $W(f)$ is a unitary. The Weyl-algebra $\A$ can be endowed with a norm and completed to a $C^*$-algebra \cite{BHR2004}. We can formally interpret $W(f)$ as $e^{i\phi(f)}$, where $\phi(f)=\int \phi(x)f(x)\mathrm{dvol}_g$ is the averaged quantum field. The relations above then express the reality of $\phi$, the equation of motion and the canonical commutation relations, respectively.

A state of the quantum field is a linear functional $\omega:\A\to\C$ which is positive, $\omega(A^*A)\ge0$, and normalized, $\omega(1)=1$. Given any state we can construct its GNS-representation, which consists of a Hilbert space $\cH_{\omega}$, a unit vector $\Omega\in\cH_{\omega}$ and a representation $\pi_{\omega}:\A\to\mathcal{B}(\cH_{\omega})$ such that $\omega(A)=\langle\Omega,\pi_{\omega}(A)\Omega\rangle_{\cH_{\omega}}$ for all $A\in\A$ and $\pi_{\omega}(\A)\Omega\subset\cH_{\omega}$ is dense. The latter two properties uniquely determine the GNS-representation up to unitary equivalence. We call $(\pi_{\omega},\cH_{\omega},\Omega)$ the GNS-triple.

We will call a state $\omega$ \emph{regular} when it has well-defined $n$-point distributions for all $n\in\N$, i.e. there are distributions $\omega_n$ on $M^{\times n}$ such that
\begin{align}
\omega_n(f_1,\ldots,f_n)&=\left.(-i)^n\partial_{s_1}\cdots\partial_{s_n}\omega(W(s_1f_1)\cdots W(s_nf_n))\right|_{s_1=\ldots=s_n=0}\notag
\end{align}
for all $f_1,\ldots,f_n$. In the GNS-representation of a regular state, $s\mapsto \pi_{\omega}(W(sf))$ is a unitary group which is strongly continuous on the dense domain $\pi_{\omega}(\A)\cH_{\omega}$ and hence everywhere. The self-adjoint generator of this group will be denoted by $\phi_{\omega}(f)$. Because $f\mapsto\phi_{\omega}(f)$ is real linear we may extend this notation by linearity to complex test functions and interpret $\phi_{\omega}(f)$ as the smeared quantum field $\phi$ in the representation $\pi_{\omega}$.

We will call a state $\omega$ \emph{quasi-free} (or \emph{Gaussian}) when it is regular and
\begin{align}
\omega(W(f))&=e^{-\frac12\omega_2(f,f)}\,.\notag
\end{align}
Note that the two-point distribution $\omega_2\in\mathcal{D}'(M^{\times 2})$ of any state is of positive type, it is a solution to the Klein-Gordon equation in each argument and it satisfies $\omega_2(x,y)-\omega_2(y,x)=iE(x,y)$. Conversely, every distribution $\omega_2\in\mathcal{D}'(M^{\times 2})$ with these three properties is the two-point distribution of a unique quasi-free state, cf. \cite{Wald1994}.

We will call a state $\omega$ \emph{Hadamard} when it is regular and its two-point distribution has
\begin{align}
WF(\omega_2)&=C_M\notag
\end{align}
in the notation of Theorem \ref{thm:JS} in the appendix, where $WF$ denotes the smooth wave front set. That theorem then gives us the Sobolev wave front set $WF^{(s)}(\omega_2)$ for all $s\in\R$. We note that for any Hadamard state $\omega$ any normalized vector in the dense linear space
$\pi_{\omega}(\A)\Omega$ in $\cH_{\omega}$ again defines a Hadamard state (cf. \cite{FV2002} Thm.4.2 and Thm.A.1 and its proof). Furthermore, the two-point distributions of any two Hadamard states differ by a smooth function.

Using Hadamard states one can define a local and covariant renormalization procedure for renormalized and time-ordered products of quantum fields, cf. \cite{HW2015} for a detailed discussion. In particular one can define a Wick square $:\phi^2:(x)$ and a renormalized stress tensor $T^{\mathrm{ren}}_{\mu\nu}(x)$. For our purposes it will be sufficient to know that these are quantum fields with the following properties. For any two Hadamard states $\omega,\bar{\omega}$ on $M$ we have
\begin{align}
\omega(:\phi^2:(x))-\bar{\omega}(:\phi^2:(x))&=\lim_{x'\to x}(\omega_2-\bar{\omega}_2)(x,x')\label{eq:TWick}\\
\omega(T^{\mathrm{ren}}_{\mu\nu}(x))-\bar{\omega}(T^{\mathrm{ren}}_{\mu\nu}(x))&=
\lim_{x'\to x}D^{\mathrm{split}}_{\mu\nu}(\omega_2-\bar{\omega}_2)(x,x')\,,\label{eq:Tquantum}
\end{align}
where the operator $D^{\mathrm{split}}_{\mu\nu}$ is defined near the diagonal in $M^{\times 2}$ by
\begin{align}
D^{\mathrm{split}}_{\mu\nu}&:=\nabla_{\mu}\otimes\nabla_{\nu}-\frac12g'_{\mu\nu}((g')^{\rho\sigma}\nabla_{\rho}\otimes\nabla_{\sigma}-m^2)\,.\notag
\end{align}
Here $g^{\rho\mu}(x)g'_{\mu\nu}(x,x')$ is the parallel propagator from $T_{x'}M$ to $T_xM$ along a geodesic between these points. For $x'$ and $x$ in a sufficiently small neighbourhood of the diagonal there is a unique geodesic in that neighbourhood, which allows us to define $D^{\mathrm{split}}_{\mu\nu}$ unambiguously near the diagonal, cf. \cite{Moretti2021}. Furthermore, in the GNS-representation of a Hadamard state $\omega$, the smeared operators $:\phi^2:(f)$ and $T^{\mathrm{ren}}_{\mu\nu}(h^{\mu\nu})$, with $f$ and $h$ compactly supported and smooth, define quadratic forms on the dense domain $\pi_{\omega}(\A)\Omega$ in $\cH_{\omega}$ (cf. \cite{BF2000}).

We are now ready to establish a new QEI, which is our main result.
\begin{theorem}\label{thm:QEI}
Let $t$ be a smooth, timelike vector field on $M$ and $f,F\in C_0^{\infty}(M,\R)$ such that $F\equiv 1$ on $\mathrm{supp}(f)$. Then there are constants $c,C>0$ such that for all Hadamard states $\omega$
\begin{align}
\omega(\phi(f)^2)&\le C(\omega(T^{\mathrm{ren}}_{\mu\nu}(t^{\mu}t^{\nu}F^2))+c)\,.\notag
\end{align}
\end{theorem}

For $f=0$ the result reduces to a known QEI \cite{FS2008}.

\begin{proof*}
We choose $s\in\R$ large enough that $WF^{(-s)}(\omega_2)=\emptyset$ for all Hadamard states, i.e. $s>\frac{d-3}{2}$ when $d$ is even and $s>\frac{d-2}{2}$ when $d$ is odd, cf. Theorem \ref{thm:JS}. We then let $C>0$ and the distribution $u$ on $M^{\times 2}$ be as in Lemma \ref{lem:positiveu} for this $s, f,t$, $\tilde{F}:=\sqrt{t_{\mu}t^{\mu}}F$ instead of $F$ and the metric volume form $\mu_g=\mathrm{dvol}_g$ on $M$. It then follows from Theorem \ref{thm:positivity} that
\begin{align}
0&\le \omega_2(Cu-f\otimes f)=C\omega_2(u)-\omega_2(f,f)\,,\notag
\end{align}
where the right-hand side is finite due to the wave front set properties of $u$ and $\omega_2$. The last term is $\omega(\phi(f)^2)$. We fix a Hadamard reference state $\bar{\omega}$, so that $c_0:=\bar{\omega}_2(u)\ge0$ by Theorem \ref{thm:positivity}. Setting $\tilde{u}(x,y):=u(y,x)$ we also have
\begin{align}
\omega_2(u)-c_0&=(\omega_2-\bar{\omega}_2)(u)=\frac12(\omega_2-\bar{\omega}_2)(u+\tilde{u})\notag\\
&=\frac12(\omega_2-\bar{\omega}_2)((\tilde{F}\otimes \tilde{F})\delta_{\mu_g})\notag\\
&=\frac12\omega(:\phi^2:(\tilde{F}^2))-\frac12\bar{\omega}(:\phi^2:(\tilde{F}^2))\notag
\end{align}
by (\ref{eq:TWick}), where we used the fact that $\omega_2-\bar{\omega}_2$ is symmetric due to the canonical commutation relations. Thus we find
\begin{align}
\omega_2(f,f)&\le C\omega_2(u)=Cc_0-\frac12C\bar{\omega}(:\phi^2:(\tilde{F}^2))+\frac12C\omega(:\phi^2:(\tilde{F}^2))\,.\notag
\end{align}
Replacing $C$ by $\frac{C}{m^2}$ we obtain
\begin{align}
\omega_2(f,f)&\le C\left(\omega\left(\frac12m^2:\phi^2:(g_{\mu\nu}t^{\mu}t^{\nu}F^2)\right)+c_1\right)\notag
\end{align}
for some $c_1>0$ such that $c_1>c_0m^2-\frac12m^2\bar{\omega}(:\phi^2:(\tilde{F}^2))$. The right hand-side involves one of the terms in the renormalized stress tensor. It is well-known that the remaining terms can be estimated from below by a constant \cite{FS2008}, leading to the final result. For convenience we reproduce the argument.

We use the notations of Lemma \ref{lem:positiveu} and assume that the local coordinate neighbourhoods $\{O_j,\kappa_j\}$ are contractible. For each $j$ we choose an orthonormal frame $\{e_{j,a}^{\mu}\}_{a=1}^d$ for the tangent bundle $TO_j$ such that $e_{j,1}^{\mu}=(t_{\nu}t^{\nu})^{-\frac12}t^{\mu}$. We then define a distribution $w$ on $M^{\times 2}$ by
\begin{align}
w(x,y)&:=(2\pi)^{-d}\int\mathrm{d}^dk\sum_{j=1}^n\sum_{a=1}^d e_{j,a}^{\mu}(x)\partial_{x^{\mu}}\tilde{F}(x)\chi_j(x) \mu_j(x)^{-\frac12} e^{i\kappa_j(x)\cdot k}\notag\\
&\quad e_{j,a}^{\mu}(y)\partial_{y^{\mu}}\tilde{F}(y)\chi_j(y)\mu_j(y)^{-\frac12}e^{-i\kappa_j(y)\cdot k}\hat{v}(k_1)\,.\notag
\end{align}
In our choice of $v$ (cf. Lemma \ref{lem:v0}) we choose $l>\frac{s}{2}+\frac{d}{4}+1$ here, where the $+1$ allows us to conclude that $WF^{(s+2)}(w)\subset \{((x,y),(p,q))\in T^*M^{\times 2}\mid p(t)\le0\le q(t)\}$ as in Lemma \ref{lem:positiveu}, despite the presence of derivatives. $w$ is compactly supported and of positive type and the wave front set estimate entails that $c_2:=\bar{\omega}_2(w)\ge0$ is finite. Furthermore, writing
$\tilde{w}(x,y):=w(y,x)$ and using Theorem \ref{thm:positivity} and (\ref{eq:Tquantum}),
\begin{align}
0&\le \omega_2(w)=c_2+(\omega_2-\bar{\omega}_2)(w)=c_2+\frac12(\omega_2-\bar{\omega}_2)(w+\tilde{w})\notag\\
&=c_2+\omega\left(T^{\mathrm{ren}}_{\mu\nu}(t^{\mu}t^{\nu}F^2)-\frac12m^2:\phi^2:(g_{\mu\nu}t^{\mu}t^{\nu}F^2)\right)
-\bar{\omega}\left(T^{\mathrm{ren}}_{\mu\nu}(t^{\mu}t^{\nu}F^2)-\frac12m^2:\phi^2:(g_{\mu\nu}t^{\mu}t^{\nu}F^2)\right)\,.\notag
\end{align}
Adding this inequality $C$ times to the right-hand side of the previous estimate and increasing $c_1$ if necessary we find we find the desired estimate.
\end{proof*}

\begin{remark}
It is clear from the proof of Theorem \ref{thm:QEI} that the result extends to adiabatic Hadamard states $\omega$ of sufficiently high order that the stress tensor is well-defined, because such states satisfy $WF^{(s)}(\omega_2-\bar{\omega}_2)=\emptyset$ for some sufficiently large $s$, cf. \cite{JS2002}.
\end{remark}

Now let $\omega$ be any Hadamard state of the free scalar field. There is a quadratic form $(T^{\mathrm{ren}}_{\omega})_{\mu\nu}(t^{\mu}t^{\nu}F^2)$ on the GNS-representation space $\cH_{\omega}$ such that
\begin{align}
\omega'(T^{\mathrm{ren}}_{\mu\nu}(t^{\mu}t^{\nu}F^2))&=\langle \psi,(T^{\mathrm{ren}}_{\omega})_{\mu\nu}(t^{\mu}t^{\nu}F^2)\psi\rangle_{\cH}\notag
\end{align}
if $\psi\in\pi_{\omega}(\A)\Omega$ is a unit vector and $\omega'$ is the Hadamard state defined by $\psi$. By Theorem \ref{thm:QEI} with $f=0$ it is known that the quadratic form $(T^{\mathrm{ren}}_{\omega})_{\mu\nu}(t^{\mu}t^{\nu}F^2)$ is bounded from below, so it has a self-adjoint Friedrichs extension,
which we will denote by $(\overline{T^{\mathrm{ren}}_{\omega}})_{\mu\nu}(t^{\mu}t^{\nu}F^2)$.\footnote{It is known that one can define $(T^{\mathrm{ren}}_{\omega})_{\mu\nu}(t^{\mu}t^{\nu}F^2)$ as an operator, but it is unknown whether this operator is essentially self-adjoint on the domain $\pi_{\omega}(\A)\Omega$. See \cite{San2012} and Thm.5.2 in \cite{San2013} for related essential self-adjointness results when $\omega$ is quasi-free.}
Theorem \ref{thm:QEI} then implies
\begin{align}
\langle \psi,\phi_{\omega}(f)^2\psi\rangle_{\mathcal{H}_{\omega}}&\le 
C\langle\psi,\left((\overline{T^{\mathrm{en}}_{\omega}})_{\mu\nu}(t^{\mu}t^{\nu}F^2)+c\right)\psi\rangle_{\mathcal{H}_{\omega}}\notag
\end{align}
for all $\psi\in\pi_{\omega}(\A)\Omega$. Because $\pi_{\omega}(\A)\Omega$ is a form core for
$(\overline{T^{\mathrm{ren}}_{\omega}})_{\mu\nu}(t^{\mu}t^{\nu}F^2)$ it is also a core for the operator
$((\overline{T^{\mathrm{ren}}_{\omega}})_{\mu\nu}(t^{\mu}t^{\nu}F^2)+c)^{\frac12}$. This leads to the operator inequality
\begin{align}
\left((\overline{T^{\mathrm{en}}_{\omega}})_{\mu\nu}(t^{\mu}t^{\nu}F^2)+c\right)^{-\frac12}
\phi_{\omega}(f)^2
\left((\overline{T^{\mathrm{en}}_{\omega}})_{\mu\nu}(t^{\mu}t^{\nu}F^2)+c\right)^{-\frac12}
&\le C\,,\label{eq:Tbound}
\end{align}
which entails $\|\phi_{\omega}(f)\left((\overline{T^{\mathrm{en}}_{\omega}})_{\mu\nu}(t^{\mu}t^{\nu}F^2)+c\right)^{-\frac12}\|\le\sqrt{C}$. This is the desired stress tensor bound on the smeared quantum field. It is interesting to note that the constants $c$ and $C$ are independent of the Hadamard state chosen, so they apply equally to all GNS-representations obtained in this way.

\begin{remark}
We consider the Rindler spacetime, $M=\{x\in\R^d\mid x^1>|x^0|\}$ as a subset of Minkowski space $M_0=(\R^d,\eta)$ in inertial coordinates $x^{\mu}$.
We denote the Weyl-algebras for the massive free scalar field on the Rindler and Minkowski spacetimes by $\A_M$ and $\A_0$, respectively, and note that
$\A_M\subset\A_0$. If $\omega_0$ denotes be Minkowski vacuum state with GNS-triple $(\pi_0,\cH_0,\Omega_0)$, then $\omega_0$ restricts to a (regular, quasi-free) Hadamard state $\omega$ state on $\A_M$ whose GNS-triple is $(\pi,\cH,\Omega)=(\pi_0|_{\A_M},\cH_0,\Omega_0)$.

There is a self-adjoint Hamiltonian operator $H_0$ in $\cH_0$ which implements the inertial time flow and for a smeared field $\phi(f)$ there is an $H$-bound of the form $\|\phi_0(f)(H_0+1)^{-\frac12}\|<\infty$ (because $H\ge mN$, where $N$ is the number operator on the Fock space $\cH_0$, cf. \cite{BW1992} Sec.12.5.4.). Similarly, $M$ is static w.r.t. the time flow of the uniformly accelerated observers and there is a self-adjoint Hamiltonian $H$ in $\cH$ which implements this time flow. Clearly $H\not=H_0$. Indeed, it is known that $\omega$ is a thermal (i.e. KMS) state w.r.t. the time flow \cite{BW1975} and hence the spectrum of $H$ is unbounded from below as well as from above. An $H$-bound of the same form, involving negative powers of $H+1$ or even $H+c$ for some positive $c$, cannot even be properly formulated.

Nevertheless, the stress tensor bound (\ref{eq:Tbound}) holds on $M$ and we are free to choose for $t^{\mu}$ either the inertial time flow, the flow of the accelerated observers, or any other smooth timelike vector field.
\end{remark}

\section{Bounds on pointwise free scalar fields in $1+1$ dimensions}\label{sec:pointwisebound}

We now turn to stress tensor bounds on the pointwise quantum field $\phi(x)$. It is clear that when the test function $f$ in Theorem \ref{thm:QEI} approaches a Dirac delta distribution, then the constant $c$ must diverge, because $\omega_2$ is divergent on the diagonal. However, we can obtain the following elementary estimate on the one-point distribution.

\begin{proposition}\label{prop:QEI1pt}
Let $(M,g)$ be a globally hyperbolic Lorentzian manifold, $t^{\mu}$ a smooth timelike vector field and $F\in C_0^{\infty}(M,\R)$. Then there is a $c>0$ such that
\begin{align}
\omega(T^{\mathrm{ren}}_{\mu\nu}(t^{\mu}t^{\nu}F^2))&\ge -c+\int_M T_{\mu\nu}[\omega_1]t^{\mu}t^{\nu}F^2\mathrm{dvol}_g\,,\notag
\end{align}
for all Hadamard states $\omega$ of a minimally coupled free scalar field, where the right-hand side integrates the classical stress tensor of the classical field configuration $\omega_1$.
\end{proposition}
\begin{proof*}
$\omega_1$ is a real-valued solution to the Klein-Gordon equation, which is smooth due to the Hadamard condition, cf. Prop.4.1 of \cite{San2010}. The truncated two-point distribution $\omega_2^T(x,y):=\omega_2(x,y)-\omega_1(x)\omega_1(y)$ has the same singularity structure as $\omega_2$ and it is again a solution to the Klein-Gordon equation in each argument. Moreover, $\omega_2^T$ is of positive type, because
\begin{align}
\omega_2^T(\bar{f},f)&=\omega_2(\bar{f},f)-|\omega_1(f)|^2
=\langle\phi_{\omega}(f)\Omega,\phi_{\omega}(f)\Omega\rangle-|\langle\Omega,\phi_{\omega}(f)\Omega\rangle|^2\ge 0\notag
\end{align}
using the Cauchy-Schwarz inequality in the GNS-representation. Thus, $\omega_2^T$ itself is a Hadamard two-point distribution and it defines a quasi-free quantum state, which we will denote by $\omega^T$.

Comparing the expectation values of the stress tensor we immediately find from (\ref{eq:Tquantum}) and (\ref{eq:Tclassical}) that
\begin{align}
\omega(T^{\mathrm{ren}}_{\mu\nu}t^{\mu}t^{\nu}F^2)-\omega^T(T^{\mathrm{ren}}_{\mu\nu}t^{\mu}t^{\nu}F^2)&=
\int_M T_{\mu\nu}[\omega_1]t^{\mu}t^{\nu}F^2\mathrm{dvol}_g\,.\notag
\end{align}
Applying the QEI of Theorem \ref{thm:QEI} with $f=0$ to the term with $\omega^T$ yields the result.
\end{proof*}

Proposition \ref{prop:QEI1pt} is a QEI which, roughly speaking, compares the quantum energy density with the classical one. This is interesting, because the classical stress tensor plays in important role in the study of classical solutions to semi-linear wave equations through the use of energy estimates.

In $1+1$ dimensions we can exploit Proposition \ref{prop:QEI1pt} to estimate pointwise quantum fields.
\begin{theorem}\label{thm:pointwiseQEI}
For every $x\in M$ and $F\in C_0^{\infty}(M)$ such that $F\equiv 1$ on an open neighbourhood of $x$ there are $c,C>0$ such that
\begin{align}
|\omega_1(x)|&\le C(\omega(T^{\mathrm{ren}}_{\mu\nu}(t^{\mu}t^{\nu}F^2))+c)\,.\notag
\end{align}
\end{theorem}
\begin{proof*}
$\omega_1$ is a real-valued solution to the Klein-Gordon equation and it is smooth due to the Hadamard condition. We can choose a chart $(O,\kappa=(x^0,\ldots,x^n))$ on a (small) open neighbourhood $O$ of $x$ such that $t^{\mu}$ is the coordinate derivative w.r.t. $x^0$, $F\equiv 1$ on $O$, $x=0$ and 
$g_{\mu\nu}(0)=\alpha\eta_{\mu\nu}$ for some $\alpha>0$. For $R>0$ we define the open sets
\begin{align}
W&:=\left\{|x^0|< R,\sum_{i=1}^n(x^i)^2<9R^2\right\}\notag\\
V_{\pm}&:=\left\{x\mid \pm x^0\in(0,R), \sum_{i=1}^n(x^i)^2<(R+2|x^0|)^2\right\}\notag
\end{align}
in $\R^d$. We may choose $R$ small enough that $\overline{W}\subset O$ and the boundary of each $V_{\pm}$ consists of three smooth, spacelike surfaces. For $t<0$ we let $V_t=V_-\cap\{x^0=t\}$ (so $V_t=\emptyset$ when $t\le -R$) and we recall the standard energy estimate
\begin{align}
\int_{V_0} T_{00}[\omega_1]\mathrm{dvol}_{g_0}&\le C_0\int_{V_t} T_{00}[\omega_1]\mathrm{dvol}_{g_t}\,,\notag
\end{align}
where $g_t$ is the Riemannian metric induced on $V_t$ and $C_0>0$ is independent of $\omega_1$ and $t\in(-R,0)$, cf. \cite{Cho2009} App.III, especially Eqn.(3.15). The analogous estimate holds for $t>0$ with $V_t=V_+\cap\{x^0=t\}$. Because the integrand on the right-hand side is non-negative it follows that
\begin{align}
\int_W T_{00}[\omega_1]\mathrm{dvol}_g&\ge C_1\int_{-R}^R\int_{V_t} T_{00}[\omega_1]\mathrm{dvol}_{g_t}\mathrm{d}t
\ge C_2\int_{V_0} T_{00}[\omega_1]\mathrm{dvol}_{g_0}\notag
\end{align}
for some $C_1,C_2>0$. Since $T_{00}[\omega_1]=T_{\mu\nu}(t^{\mu}t^{\nu}F^2)$ on $W$ we find from Proposition \ref{prop:QEI1pt} that
\begin{align}
\omega(T^{\mathrm{ren}}_{\mu\nu}(t^{\mu}t^{\nu}F^2))&\ge -c+C_2\int_{V_0} T_{00}[\omega_1]\mathrm{dvol}_{g_0}\notag
\end{align}
for some $c>0$.

Note that $\omega_1$ and its derivatives restricted to the ball $V_0$ are (in local coordinates) all in $L^2$. If $d=1+1$ we can use a Morrey's Inequality, \cite{Eva2010} Sec.5.6 Thm.4, to find
\begin{align}
|\omega_1(x)|^2&\le C_4\int_{V_0} T_{00}[\omega_1]\mathrm{dvol}_{g_0}\le C(\omega(T^{\mathrm{ren}}_{\mu\nu}(t^{\mu}t^{\nu}F^2))+c)\,.\notag
\end{align}
Increasing $c$ by $C^{-1}$ if necessary we find that the right-hand side is less than or equal to its own square, so taking square roots yields the result.
\end{proof*}

If $\omega$ is any (not necessarily quasi-free) Hadamard state of the massive minimally coupled scalar field in a $1+1$ dimensional Lorentzian manifold, then Theorem \ref{thm:pointwiseQEI} implies
\begin{align}
|\langle \psi,\phi_{\omega}(x)\psi\rangle_{\cH_{\omega}}|&\le C\langle \psi,((\overline{T^{\mathrm{ren}}_{\omega}})_{\mu\nu}(t^{\mu}t^{\nu}F^2)+c)\psi\rangle_{\cH_{\omega}}\notag
\end{align}
for all $\psi\in\pi_{\omega}(\A)\Omega$, where we use the notations of Section \ref{sec:smearedbounds}. Consequently,
\begin{align}
-C&\le((\overline{T^{\mathrm{ren}}_{\omega}})_{\mu\nu}(t^{\mu}t^{\nu}F^2)+c)^{-\frac12}\phi_{\omega}(x)
((\overline{T^{\mathrm{ren}}_{\omega}})_{\mu\nu}(t^{\mu}t^{\nu}F^2)+c)^{-\frac12}\le C\label{eq:pointwiseTbound}
\end{align}
for suitable $c,C>0$. This is the desired stress tensor bound for pointlike fields. Once again it is interesting to note that $c$ and $C$ are independent of $\omega$.

\section{Discussion}\label{sec:discussion}

In Equations (\ref{eq:Tbound}) and (\ref{eq:pointwiseTbound}) we have shown that the unboundedness of smeared quantum field operators and the singularity of pointwise quantum field operators in $1+1$ dimensions can be controlled by polynomials in an averaged quantum stress tensor. This is true in general globally hyperbolic Lorentzian manifolds, which is a significant improvement over pre-existing $H$-bounds that necessarily require a ground state representation and hence a static metric. Before we discuss the possible implications of these stress tensor bounds for more general (possibly interacting) theories, let us first point out some of their general features.

Firstly, Theorems \ref{thm:QEI} and \ref{thm:pointwiseQEI}, from which the stress tensor bounds are derived, are formulated as new types of QEIs. Whereas QEIs have previously mostly been investigated with a view to applications in GR \cite{Few2017,KS2020} (e.g. singularity theorems for quantum matter) or the stability of matter \cite{FV2002}, these new QEIs change the perspective and provide information about the structure of the quantum field theory itself. To prove these QEIs we used an apparently new result on distributions of positive type, Theorem \ref{thm:positivity}.

Secondly, the reader will have observed that the new aspects of Theorem \ref{thm:QEI} are essentially about a quantum inequality for the Wick square, which appears in the stress tensor when $m>0$. If the claim remains true in the massless case, the proof will need substantial modification. Similarly, \ref{thm:pointwiseQEI} relies heavily on the low dimension. We conjecture that a similar result involving higher powers of the stress tensor is valid in higher dimensions, but it will require a substantially modified proof involving higher $n$-point distributions.

Thirdly, the stress tensor bound (\ref{eq:Tbound}) on smeared quantum fields requires the same power of the stress tensor, regardless of the dimension.
For chiral CFTs in $1+1$ dimensions, however, the analogous local energy bounds of \cite{CTW2022} may involve different powers of the energy density,
depending on the scaling dimension of the field. This suggests that stress tensor bounds for Wick polynomials and other smeared quantum fields may also
require different powers of the stress tensor, depending on their scaling degree.

Finally, the stress tensor bounds that we found involve constants $c,C>0$ that are independent of the representation, as long as we consider GNS-representations of Hadamard states. This suggests that the bounds tell us about some intrinsic properties of the theory and they could potentially be used to define a topology on the Weyl algebra (in a local and covariant way), which is weaker than the $C^*$-norm topology and which allows us to recover smeared field operators by taking a completion. This idea is similar to the H\"ormander topology that can be used to extend the $^*$-algebra of smeared fields in order to include e.g. Wick polynomials \cite{BF2000,DB2014}. However, there are two key differences. Firstly, the analytic structure we envisage here targets bounded operators and therefore connects to the good spectral properties that such operators have. Secondly, there may be advantages to using the stress tensor to define such a topology instead of resorting to the H\"ormander topology. These ideas will be the subject of a future investigation.

If one takes the point of view that a good quantum field theory should be local, covariant and have a quantum stress tensor, then one has all the ingredients necessary to formulate QEIs like those of Theorems \ref{thm:QEI} and \ref{thm:pointwiseQEI} and the ensuing stress tensor bounds (\ref{eq:Tbound}) and (\ref{eq:pointwiseTbound}). This would open up a way to connect the perspective of bounded operator algebras and their spectral theory \cite{BFV2003} to the perspective of field operators with their geometric meaning and differential equations \cite{HW2010}. So far, the connection between these perspectives is only well-understood in Minkowski space, cf. \cite{Bos2005b,Bos2005a,BW1992}. The
main obstruction to generalising results about this connection to general Lorentzian manifolds is the need to have a Hamiltonian operator. Our results indicate that this obstruction can potentially be overcome by using stress tensor bounds.

\medskip

\textbf{Acknowledgements} I thank Rainer Verch for encouraging comments and for suggesting the reference \cite{BW1992} and Sebastiano Carpi for pointing out connections to chiral CFTs in $1+1$ dimensions. I am grateful to two anonymous reviewers for detailed comments and suggestions. I also thank the participants and organisers of the conference Energy conditions in quantum field theory in Leipzig (2022), where initial results of this project were presented.

\appendix

\section{Sobolev wave front sets and the Hadamard parametrix}\label{app:WFs}

In this appendix we recall the definition of Sobolev wave front sets and give a few results that will be needed in the main text. For further properties and results we refer to Appendix B of \cite{JS2002}.

We consider a smooth orientable manifold $M$ of dimension $d\in\N$ with a smooth non-vanishing volume form $\mu$. We will freely identify smooth functions $u\in C^{\infty}(M)$ with distributions in $\mathcal{D}'(M)$ using $u(f)=\int_Muf\mu$ for all $f\in C_0^{\infty}(M)$.

For any $u\in\mathcal{D}'(M)$, $s\in\R$ and $(x,\xi)\in T^*M$ with $\xi\not=0$ we will write $u\in H^{(s)}(x,\xi)$ iff there is a coordinate chart
$\kappa:O\to\R^d$ with $x\in O$, a $f\in C_0^{\infty}(O)$ with $f(x)\not=0$ and an open convex cone $\Gamma\subset\R^d$ with
$(D_{\kappa(x)}\kappa^{-1})^t\xi\in\Gamma$ such that
\begin{align}
\int_{\Gamma}\mathrm{d}^dk(1+|k|^2)^s|\widehat{(fu)\circ\kappa^{-1}}(k)|^2&<\infty\,.\notag
\end{align}
If this inequality holds, it still holds when we replace $f$ by $fh$ for any $h\in C^{\infty}(M)$. Furthermore, the definition of $H^{(s)}(x,\xi)$ can be shown to be independent of the choice of local chart $\kappa$. The Sobolev wave front set is then defined as
\begin{align}
WF^{(s)}(u)&=\{(x,\xi)\in T^*M|\ \xi\not=0,\ u\not\in H^{(s)}(x,\xi)\}\,.\notag
\end{align}
By definition $WF^{(s)}(u)$ is a closed subset (possibly empty) of $T^*M\setminus\mathcal{Z}$, where $\mathcal{Z}$ is the zero section of $T^*M$. The smooth wave front set of a distribution $u$ can be defined as the closure $WF(u)=\overline{\bigcup_{s\in\R}FW^{(s)}(u)}$ in $T^*M\setminus\mathcal{Z}$.

The following lemma is a special case of Prop.B.5 in \cite{JS2002}, where we note that $WF^{(t)}(1)=\emptyset$ for all $t\in\R$.
\begin{lemma}\label{lem:tensorWFs}
For any two smooth orientable manifolds $X$ and $Y$ with smooth non-vanishing volume forms $\mu_X$ and $\mu_Y$, respectively, and any distribution
$u\in\mathcal{D}'(X)$, the distribution $u\otimes 1$ on $X\times Y$ (with volume form $\mu_X\otimes\mu_Y$) has
\begin{align}
WF^{(s)}(u\otimes 1)&\subseteq WF^{(s)}(u)\times(Y\times\{0\})\,.\notag
\end{align}
\end{lemma}

We will also need the following variation of this result.

\begin{lemma}\label{lem:v}
For $d\in\N$, $d\ge2$, let $v_d\in\mathcal{D}'(\R^d)$ such that $v_d=v\otimes\delta$ with $v\in\mathcal{D}'(\R)$ and $\delta\in\mathcal{D}'(\R^{d-1})$. Then
\begin{align}
WF^{(s)}(v_d)\cap\{(x,k)\in\R^{2d}\mid k_1\not=0\}&=WF^{(s+\frac12(d-1))}(v)\times(\{0\}\times\R^{d-1})\,.\notag
\end{align}
\end{lemma}
\begin{proof*}
Writing $x=(x_1,\tilde{x})$ we note that $v_d$ vanishes outside $\mathrm{supp}(v)\times\{0\}$, so it is clearly smooth there. We now fix
$x=(x_1,0)\in\mathrm{supp}(v)\times\{0\}$ and choose $f\in C_0^{\infty}(\R^d)$ with $f(x)=1$. We let $f_1(x_1):=f(x_1,0)$, so that
$\widehat{fv_d}(k)=\widehat{f_1v}(k_1)$.

Let $p=(p_1,\tilde{p})\in\R^d$ with $p_1\not=0$. We can choose $\alpha\not=0$ such that $\alpha p_1>|\tilde{p}|$ and then define the open cone
$\Gamma=\{k=(k_1,\tilde{k})\in\R^d\mid \alpha k_1>|\tilde{k}|\}$, so that $p\in\Gamma$. For $k\in\Gamma$ we have
$1+k_1^2\le 1+|k|^2\le (1+\alpha^2)(1+k_1^2)$ and hence the convergence of
\begin{align}
\int_{\Gamma}(1+|k|^2)^s|\widehat{fv_d}(k)|^2\mathrm{d}^dk\label{eqn:integral}
\end{align}
is equivalent to the convergence of
\begin{align}
\int_0^{\infty} (1+|k_1|^2)^s k_1^{d-1}|\widehat{f_1v}(\pm k_1)|^2\mathrm{d}k_1\notag
\end{align}
where the sign is the sign of $p_1$ and the factor $k_1^{d-1}$ comes from integrating the coordinates $\tilde{k}$ over a ball of radius $\alpha k_1$. The convergence of the latter integral is equivalent to $(x_1,p_1)\not\in WF^{(s+\frac{d-1}{2})}(v)$. This proves the lemma.
\end{proof*}

When $M$ is a globally hyperbolic Lorentzian manifold and $\mu$ the metric volume form, then we can consider the Sobolev wave front set of a Hadamard two-point distribution $\omega_2$ on $M^{\times 2}$. As in Lemma 5.2 of \cite{JS2002} one can show

\begin{theorem}\label{thm:JS}
Let $M$ be a globally hyperbolic Lorentzian manifold. Then any Hadamard two-point distribution $\omega_2\in\mathcal{D}'(M\times M)$ satisfies
\begin{align}
WF^{(s)}(\omega_2)&=\left\{
\begin{array}{ll}
\emptyset&\mathrm{if}\ d\ \mathrm{is\ even\ and}\ s<\frac{3-d}{2},\ \mathrm{or}\ d\ \mathrm{is\ odd\ and}\ s<\frac{2-d}{2}\\
C_M&\mathrm{if}\ d\ \mathrm{is\ even\ and}\ s\ge\frac{3-d}{2},\ \mathrm{or}\ d\ \mathrm{is\ odd\ and}\ s\ge\frac{2-d}{2}
\end{array}
\right.\,,\notag
\end{align}
where
\begin{align}
C_M&=\{((x',x),(\xi',\xi))\in T^*(M\times M)|\ (x',\xi')\sim(x,-\xi),\ \xi'\ \mathrm{null\ and\ future\ pointing}\}\,,\notag
\end{align}
and $(x',\xi')\sim(x,\xi)$ means that there is a geodesic $\gamma$ between $x'$ and $x$ to which $\xi'$ and $-\xi$ are cotangent and each other's parallel transports. When $x'=x$ this reduces to $\xi'=-\xi$ which must be a future pointing null vector.
\end{theorem}

\end{document}